\documentclass[12pt,twoside]{article}
\usepackage{fleqn,espcrc1}

\usepackage{graphicx}
\usepackage{epsfig}
\newcommand{\AmS}{{\protect\the\textfont2
  A\kern-.1667em\lower.5ex\hbox{M}\kern-.125emS}}

\def\lsim{\raise0.3ex\hbox{$<$\kern-0.75em\raise-1.1ex\hbox{$\sim$}}}
\def\gsim{\raise0.3ex\hbox{$>$\kern-0.75em\raise-1.1ex\hbox{$\sim$}}}
\newcommand{\ie}{{\sl i.e.~\/}}

\title{
\vskip -80pt
\mbox{} \hfill BI-TP 2001/06\\
\mbox{} \hfill March 2001\\
\vskip 25pt
Lattice Results on QCD Thermodynamics}
\author{Frithjof Karsch\thanks{The work 
has been supported by the TMR network ERBFMRX-CT-970122
and the DFG under grants Ka 1198/8-1 and KON191/2001.}
\\
\vskip 6pt
Fakult\"at f\"ur Physik, Universit\"at Bielefeld, D-33615 Bielefeld, Germany
}      
\begin{document}

\maketitle

\begin{abstract}
We review recent results on QCD at finite temperature. Main
emphasis is put on a discussion of observables which are of immediate
interest to experimental searches for the Quark Gluon Plasma, \ie
the phase transition temperature, the equation of state in 2 and 
3-flavour QCD, the heavy quark free energy and thermal effects on 
hadron masses. 
\end{abstract}

\section{Introduction}

Since many years the thermodynamics of quarks and gluons, 
the transition between the low temperature hadronic phase 
and the quark-gluon plasma (QGP) as well as non-perturbative
properties of the high temperature phase, have been analyzed through 
numerical calculations within the framework of lattice regularized QCD.
In particular, the studies performed in the pure gauge sector
did provide a rather detailed picture of the high temperature
plasma phase as a medium of strongly interacting partons 
which are strongly screened and influenced by
non-perturbative effects even at fairly high temperatures.

Over recent years thermodynamic calculations on the lattice have 
steadily been improved. This partly is due to the much improved computer
resources, however equally important has been and still is the 
development of improved discretization schemes, \ie improved actions.
This is of particular relevance for thermodynamic calculations which
are not only sensitive to the long distance physics at the phase
transition but also probe properties at short distances in calculations 
of e.g. the energy density or the heavy quark potential.
The advantages of improved actions for thermodynamic
calculations have been discussed at recent lattice conferences
\cite{Karpisa,Eji00}, this does include improvements
of the standard Wilson and staggered fermion actions as well as
new developments of chiral actions like the domain wall and overlap
fermion formulations. We thus will not go into details of these
formulations here. 
We rather will concentrate on a discussion of 
results, which may be of direct interest to experimental
studies of the quark-gluon plasma. We will discuss
the QCD transition temperature and equation of state and will
emphasize the quark mass and flavour dependence of these quantities 
which gives some insight into the mechanisms
controlling the transition to the Quark Gluon Plasma.
Furthermore we will discuss thermal modifications of the heavy quark 
free energy and meson masses.   

\begin{figure}[htb]
\vspace{-0.6truecm}
\begin{center}
\epsfig{bbllx=59,bblly=175,bburx=564,bbury=514,
file=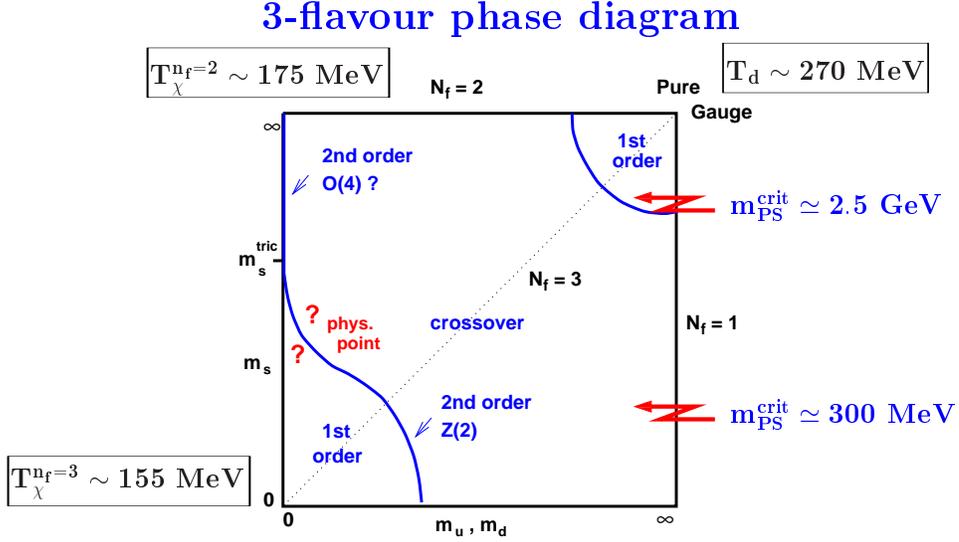,width=110mm}
\end{center}
\vskip -0.9truecm
\caption{The QCD phase diagram of 3-flavour QCD with degenerate
(u,d)-quark masses and a strange quark mass $m_s$. }
\vskip -0.8truecm
\label{fig:phased}
\end{figure}

We also refrain from discussing details of the ongoing investigations
of the order of the QCD phase transition \cite{Ber00,Ali01}, which is
first order in the case of three light, degenerate quark flavours and 
most likely is
second order in the case of 2-flavour QCD, although in this latter case
questions concerning the universality class of the transition and,
in particular,
the role of the axial anomaly for the transition are not
yet settled completely. Our current understanding of the QCD phase
diagram of 3-flavour QCD at vanishing baryon number density is shown
in Figure~\ref{fig:phased}. 

\section{The Critical Temperature and the QCD Phase Diagram} 

\begin{figure*}[t]
\vskip -0.6truecm
\vspace{9pt}
\hspace*{-0.2cm}\epsfig{file=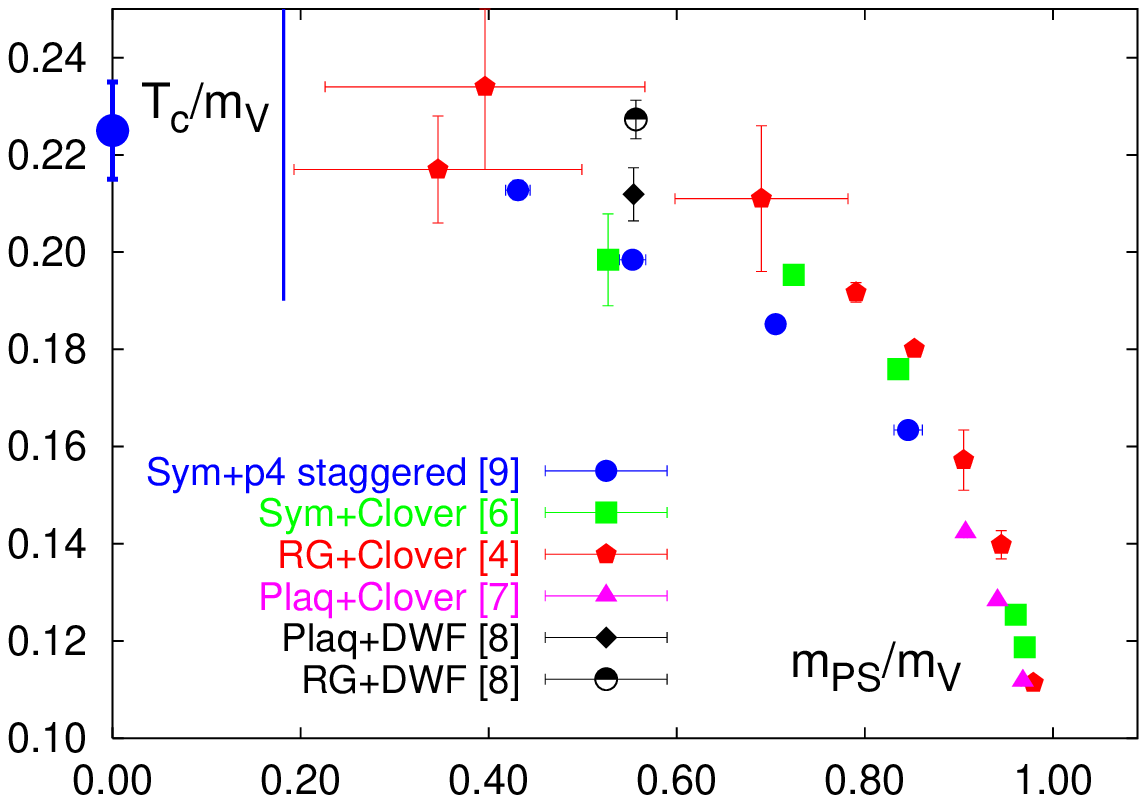,width=82mm}
\hspace*{-0.2cm}\epsfig{file=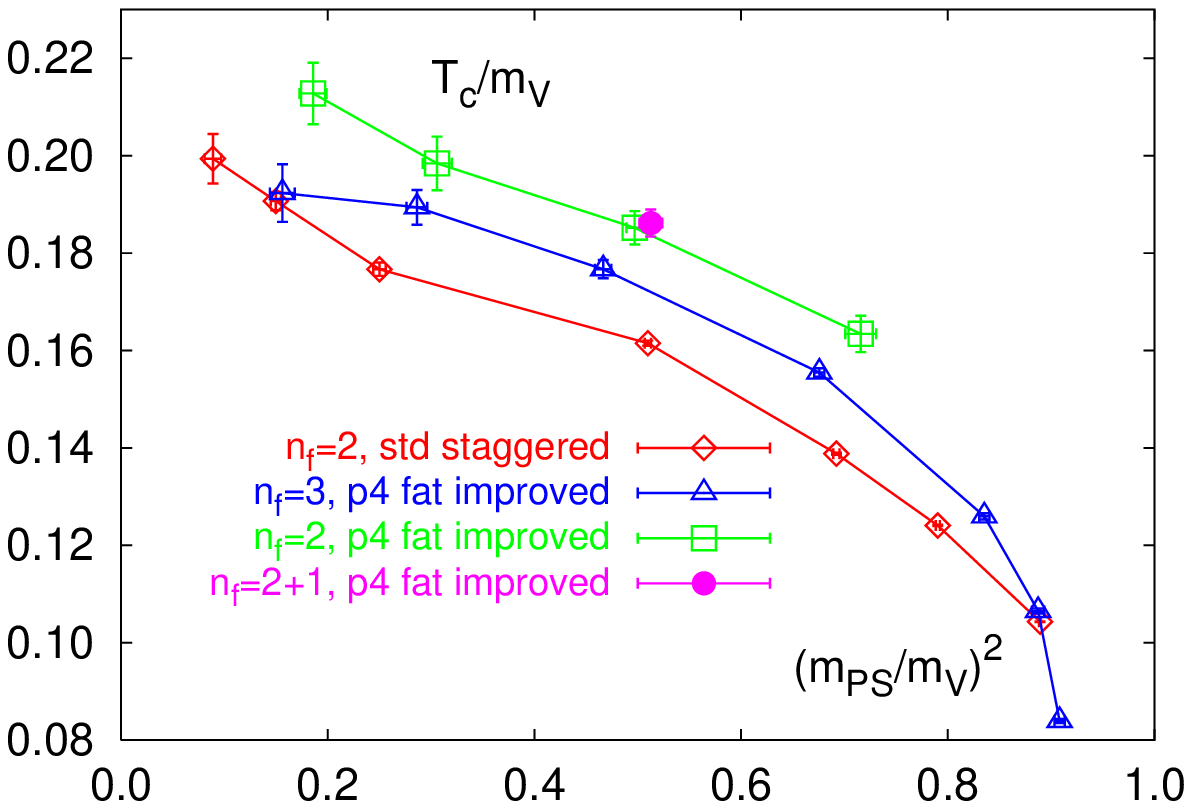,width=82mm}
\begin{picture}(17,0.1)
\put(6.7,5.2){(a)}
\put(14.5,5.2){(b)}
\end{picture}
\vskip -0.7truecm
\caption{Transition temperatures in units of $m_V$. In Fig.~2a we give
a collection of results obtained for 2-flavour QCD with various fermion
actions while Fig.~2b shows a comparison of results obtained in
2 and 3-flavour QCD with unimproved and improved staggered fermion
actions. All results are from simulations on lattices with temporal
extent $N_\tau = 4$. The large dot drawn for $m_{PS}/m_V =0$  
indicates the result of chiral extrapolations based on 
calculations with improved Wilson \cite{Ali01} as well as improved
staggered \cite{Kar00a} fermions.
The vertical line shows the location of the physical limit, 
$m_{PS}\equiv m_\pi,~m_V \equiv m_\rho $. 
}
\vskip -0.3truecm
\label{fig:tcnf2}
\end{figure*}

The early calculations of the QCD transition temperature, 
which have been performed with standard Wilson \cite{Bit91} and 
staggered \cite{Ber97} fermion actions, led to significant discrepancies 
of the results. These differences strongly diminished
in the newer calculations which are based on improved Wilson fermions
(Clover action) \cite{Ali01,Ber97,Edw99}, domain wall fermions \cite{norman} 
as well as improved staggered fermions (p4-action) \cite{Kar00a}. A 
compilation of these newer results is shown in Figure~\ref{fig:tcnf2} for
various values of the quark masses. 
To compare results obtained with different
actions the results are presented in terms of a {\it physical 
observable}, \ie the ratio of the lightest pseudo-scalar and vector 
meson masses ($m_{PS}/m_V$). In Figure~\ref{fig:tcnf2}a we
show $T_c/m_V$ obtained for 2-flavour QCD while Figure~\ref{fig:tcnf2}b
gives a comparison of results obtained with improved 
staggered fermions \cite{Kar00a} for 2 and 3-flavour QCD.
Also shown there is a result for the case of (2+1)-flavour QCD, \ie
for two light and one heavier quark flavour degree of freedom.
Unfortunately the quark masses in this latter case are still
too large to be compared directly with the situation realized in nature. 
We note however, that the results obtained so far suggest that the transition 
temperature in (2+1)-flavour QCD is close to 
that of 2-flavour QCD. The 3-flavour theory, on the other hand, leads to 
consistently smaller values of the critical temperature, 
$T_c(n_f=2)-T_c(n_f=3) \simeq 20$~MeV. Extrapolation of the transition
temperatures to the chiral limit gave
\begin{eqnarray}
\underline{\rm 2-flavour~ QCD:} &&
T_c  = \cases{(171\pm 4)\; {\rm MeV}, &  clover-improved
Wilson fermions \cite{Ali01} \cr
(173\pm 8)\; {\rm MeV}, & improved staggered fermions  \cite{Kar00a}}
\nonumber \\ 
\underline{\rm 3-flavour~ QCD:} &&
T_c  = \; \; \;
(154\pm 8)\; {\rm MeV}, \; \; \;  {\rm improved~ staggered~ fermions~}  
\cite{Kar00a}
\nonumber 
\end{eqnarray}
Here $m_\rho$ has been used to set the scale for $T_c$.
Although the agreement between results obtained with Wilson and 
staggered fermions is striking, one should bear in mind that all
these results have been obtained on lattice with temporal extent $N_\tau =4$,
\ie at rather large lattice spacing, $a\simeq 0.3$~fm.
Moreover, there are uncertainties involved in the ansatz used to
extrapolate to the chiral limit. We thus estimate that the systematic
error on the value of $T_c /m_\rho$ still is of similar magnitude as
the purely statistical error quoted above.  

We note from Figure~\ref{fig:tcnf2} that $T_c/m_V$ drops with increasing
ratio $m_{PS}/m_V$, \ie with increasing quark mass. Apparently this does 
not reflect the general picture we have for the quark mass ($m_q$) dependence 
of $T_c$. 
For instance, a simple percolation picture for the QCD transition 
would suggest that $T_c (m_q)$ or better $T_c(m_{PS})$ will increase
with increasing $m_q$; with increasing $m_q$ 
also the hadron masses increase and it becomes more difficult to 
excite the low lying hadronic states. It thus becomes more difficult to 
create a sufficiently high particle/energy density in the  
hadronic phase that can trigger a phase (percolation) transition. Such a 
picture also follows from chiral model calculations \cite{chiralmodel}. 

In order to quantify the quark mass dependence of $T_c$ we ideally 
should find an observable to set the scale for $T_c$, which itself is not
dependent on $m_q$. This clearly is not the
case for $m_V$, which actually vanishes in the limit $m_q \rightarrow \infty$.
On the other hand we know from studies of the hadron spectrum and string
tension in quenched QCD (SU(3) gauge theory) that these observables agree
with experiment and/or QCD phenomenology quite well despite the fact
that dynamical light quark contributions are suppressed in this limit. 
It thus may be expected that (partially) quenched
hadron masses or the string tension do provide an almost quark mass
independent scale. In fact, this
is what tacitly has been assumed when one converts the critical
temperature of the SU(3) gauge theory $T_c/\sqrt{\sigma} \simeq 0.63$ into
physical units\footnote{We use here and in the following 
$\sqrt{\sigma} \simeq 425~{\rm MeV}$ which 
may be deduced from quenched spectrum calculations 
($m_\rho /\sqrt{\sigma} = 1.81~(4)$ \cite{Wit97}). 
Similar values have been obtained from partially quenched calculations
in 3-flavour QCD \cite{Kar00a}.}
as has been done also in Figure~\ref{fig:phased}.   

\begin{figure}[t]
\begin{center}
\hspace*{-0.2cm}\epsfig{file=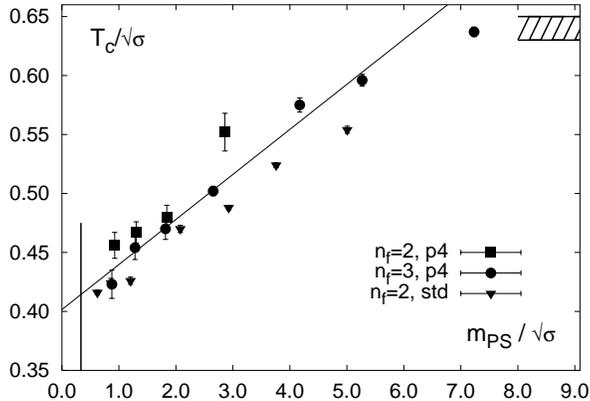,width=82mm}
\end{center}
\vskip -0.9truecm
\caption{The transition temperature in 2 (filled squares) and 3 (circles)
flavour  QCD versus $m_{PS}/\sqrt{\sigma}$ using an improved staggered
fermion action (p4-action). Also shown are results for 2-flavour QCD
obtained with the standard staggered fermion action (open squares). 
The dashed band indicates the uncertainty on $T_c/\sqrt{\sigma}$ in the 
quenched limit. The straight line is the fit given in Eq.~2.}
\vskip -0.2truecm
\label{fig:tc_pion}
\end{figure}

To quantify the quark mass dependence of the transition temperature
one may express $T_c$ in units of $\sqrt{\sigma}$.
This ratio is shown in Figure~\ref{fig:tc_pion} as a function of
$m_{PS} / \sqrt{\sigma}$. As can be seen the transition
temperature starts deviating from the quenched values for $m_{PS}\; 
\lsim \; (6-7)\sqrt{\sigma}\simeq 2.5~{\rm GeV}$. We also note that the 
dependence of $T_c$ on $m_{PS}/\sqrt{\sigma}$
is almost linear in the entire mass interval. 
Such a behaviour might, in fact, be expected
for light quarks in the vicinity of 
a $2^{nd}$ order chiral transition where the pseudo-critical temperature 
depends on the mass of the Goldstone-particle like
\begin{equation}
T_c(m_{\pi}) - T_c(0) \sim m_{\pi}^{2/\beta\delta} ~.
\end{equation} 
For 2-flavour QCD the critical indices are expected to 
belong to the universality class of 3-d, $O(4)$
symmetric spin models and one thus would indeed expect $2/\beta\delta=1.1$. 
However, this clearly cannot be the origin of the quasi linear
behaviour which is observed for rather large hadron masses and seems
to be independent of $n_f$.
Moreover, unlike in chiral models \cite{chiralmodel} the dependence
of $T_c$ on $m_{PS}$ turns out to be rather weak. The line shown in
Figure~\ref{fig:tc_pion} is a fit to the 3-flavour data, which gave
\begin{equation}
(T_c / \sqrt{\sigma} )_x = (T_c / \sqrt{\sigma} )_0 +
0.04(1)\; x \quad {\rm with} \quad x = m_{PS}/ \sqrt{\sigma} \quad .
\end{equation}

For the quark masses currently used in lattice calculations
a resonance gas model combined with a suitably chosen percolation 
criterion would probably be more appropriate to describe the 
thermodynamics close to $T_c$.

\section{The Equation of State}

When discussing the equation of state of QCD, e.g. the temperature 
dependence of the energy density ($\epsilon$) and pressure ($p$), 
we should at least 
distinguish three regimes; the high temperature regime ($T\gsim 1.5 T_c$),
the critical region ($T\simeq T_c$) and the low temperature regime
($T\lsim 0.9T_c$). The calculation of $\epsilon$ as well as $p$ on the
lattice is most difficult below $T_c$ where these observables are 
exponentially suppressed, which is true even for realistic pseudo-scalar
meson masses, $m_{PS}\sim T_c$. We therefore will concentrate on a 
discussion of the two former temperature regimes.

At high temperature we expect that $p/T^4$ and $\epsilon /T^4$ will
asymptotically approach the free gas limit for a gas of gluons and
$n_f$ quark flavours,
\begin{equation}
{\epsilon_{SB} \over T^4} = {3p_{SB} \over T^4} = 
\biggl( 16 + {21 \over 2} n_f\biggr) {\pi^2 \over 30} \quad .
\label{esb}
\end{equation} 
From calculations in the quenched limit,
the pure SU(3) gauge theory, we know that the high-T ideal gas limit
is reached only very slowly \cite{Karpisa}. In fact, for 
$T \simeq (2-4)T_c$ thermodynamic quantities deviate by
about 15\% from the limiting ideal gas value. This deviation is too
big to be understood in terms of ordinary high temperature perturbation
theory which converges badly at these low temperatures \cite{eospert}.
However, it is naturally accounted for in quasi-particle 
models \cite{quasi} and resummed perturbative calculations \cite{resummed}.

\begin{figure*}[t]
\vskip -0.8truecm
\vspace{9pt}
\hspace*{-0.2cm}\epsfig{file=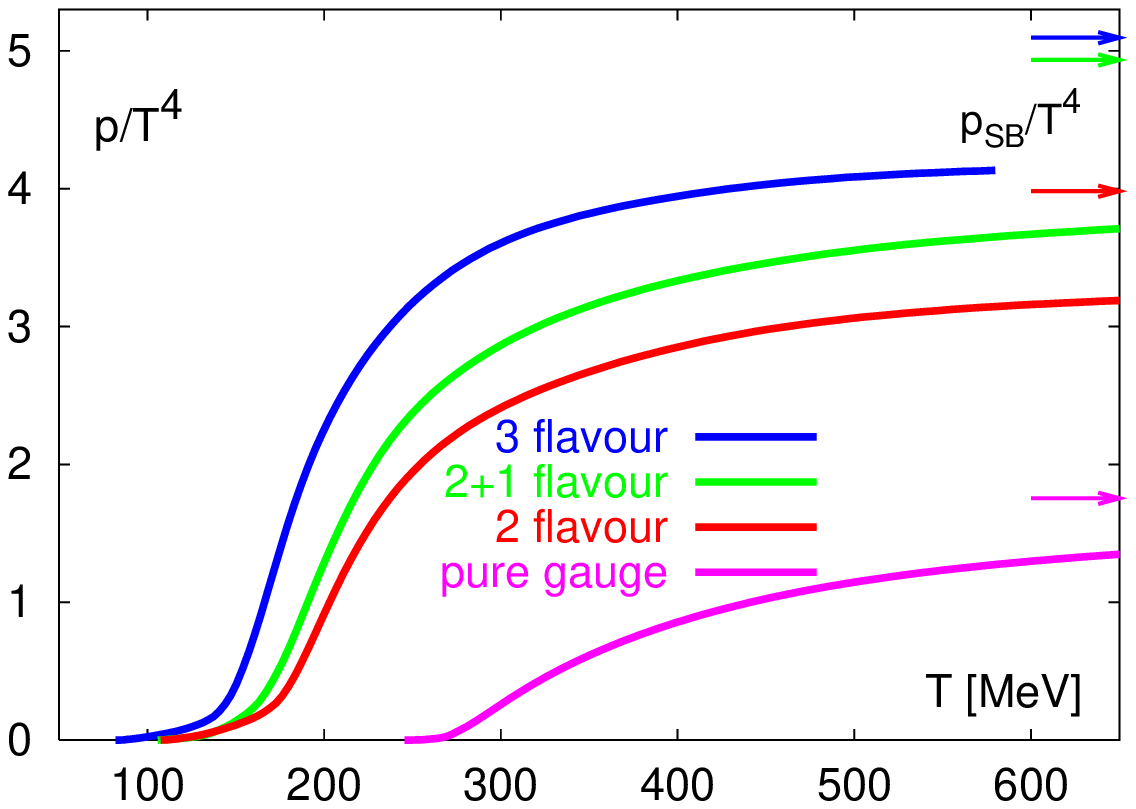,width=82mm}
\hspace*{-0.2cm}\epsfig{file=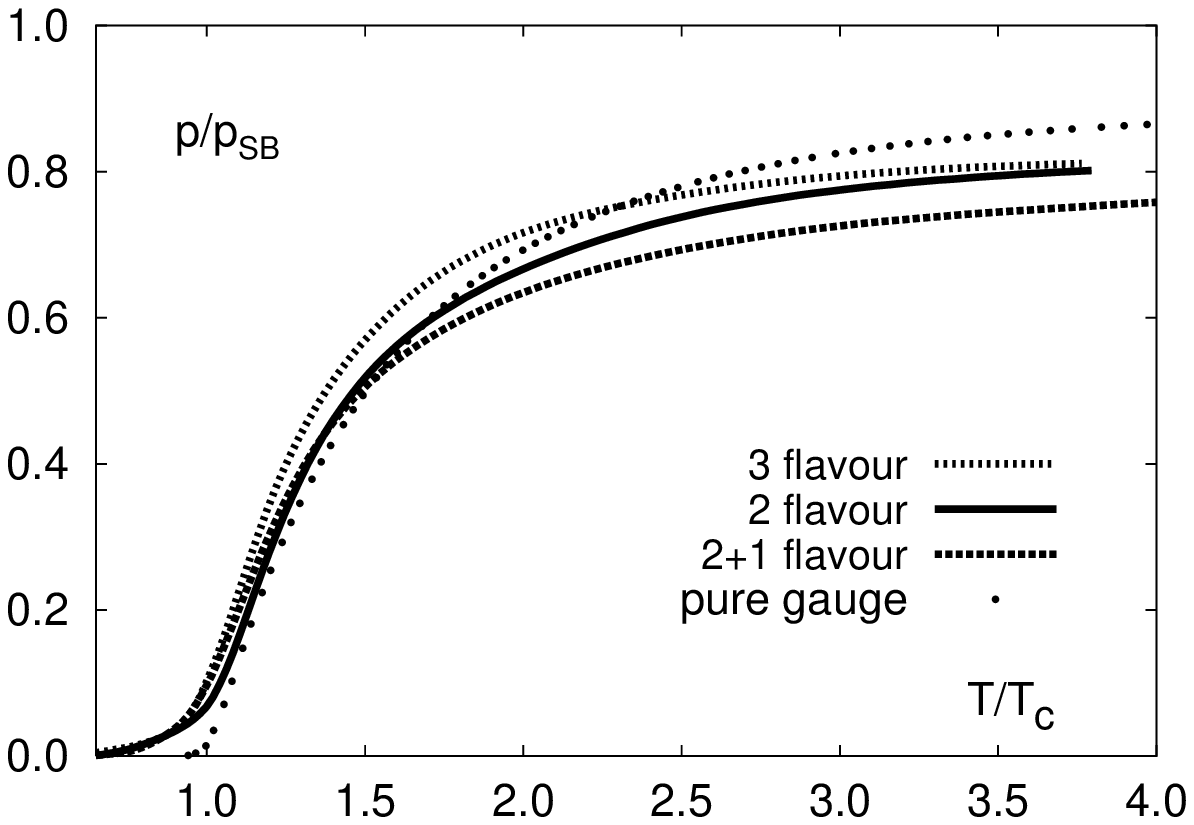,width=82mm}
\begin{picture}(17,0.1)
\put(4.0,5.2){(a)}
\put(11.8,5.2){(b)}
\end{picture}
\vskip -0.8truecm
\caption{The pressure in QCD with $n_f = 0,~2$ and 3 light quarks as
well as two light and a heavier (strange) quark. For $n_f \ne 0$ 
calculations have been performed on a $N_\tau=4$ lattice using 
improved gauge and staggered fermion actions. In the case of the SU(3)
pure gauge theory the continuum extrapolated result is shown. Arrows
indicate the ideal gas pressure $p_{SB}$ as given in Eq.~3.
}
\vspace{-0.4truecm}
\label{fig:pressure_nf2}
\end{figure*}

A similar deviation from ideal gas behaviour has now been found in 
simulations of QCD with 2 and 3 degenerate quark flavours as well as in a 
simulation with two light and one heavier quark mass \cite{Kar00}. The 
results of this calculation which has been performed with an improved 
gauge and an improved staggered fermion action (p4-action)
are shown in Figure~\ref{fig:pressure_nf2}. 
As the p4-action is known to lead to much 
smaller cut-off distortions in the high-T limit than the standard
actions, it also becomes meaningful to compare these results 
with continuum models and perturbative calculations at high 
temperature \cite{resummed}. We stress, however, that a final extrapolation 
to the continuum limit still has to be done for QCD with light quarks.
From an analysis of the cut-off dependence of the p4-action
and the experience gained in the pure gauge sector one expects that the 
results shown in Figure~\ref{fig:pressure_nf2} are still systematically
below the final continuum extrapolated result. 

The pressure shown in Figure~\ref{fig:pressure_nf2}a for QCD with different
number of flavours as well as for the pure SU(3) gauge theory clearly reflects
the strong change in the number of degrees of freedom in the high temperature
phase. Moreover, the dependence of $T_c$ on the number of partonic
degrees of freedom is clearly visible. In view of this it 
indeed is striking that $p/p_{SB}$ is almost flavour independent when
plotted in units of $T/T_c$ 
(Figure~\ref{fig:pressure_nf2}b).

\begin{figure*}[t]
\hspace*{-0.3cm}\begin{minipage}[t]{80mm}
\epsfig{file=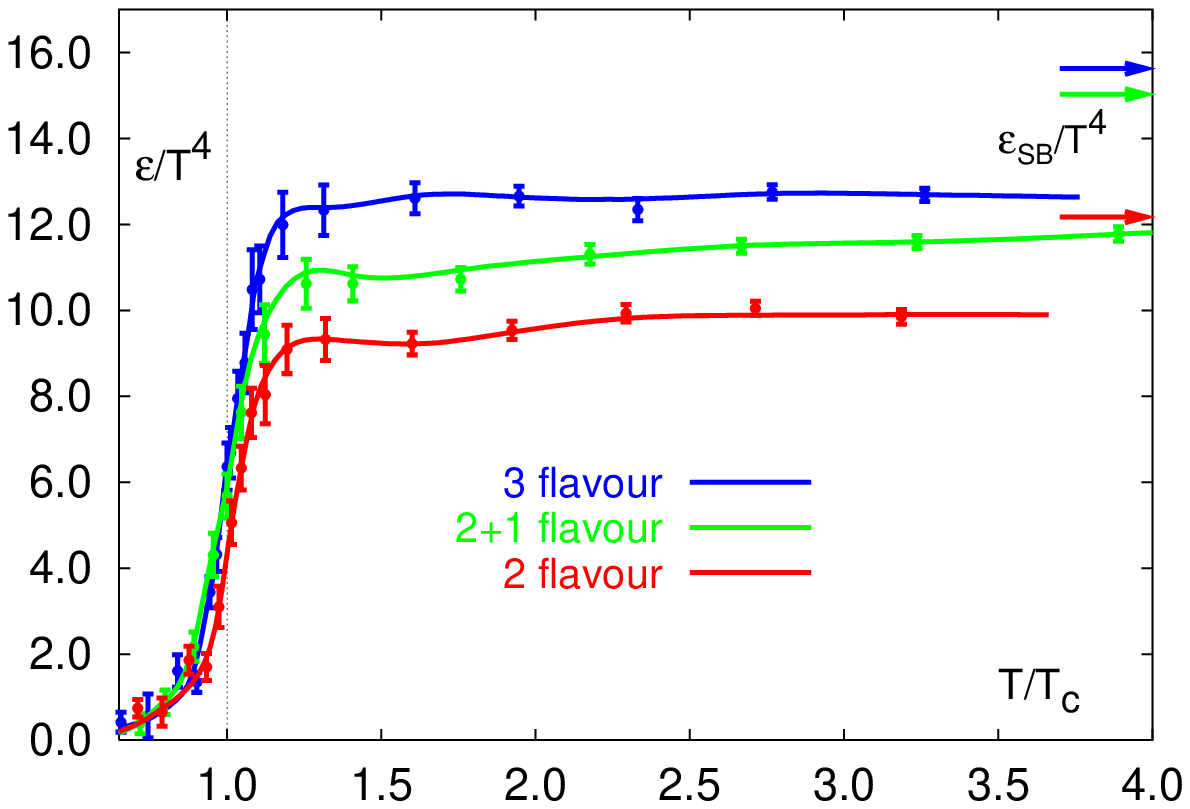,width=82mm}
\end{minipage}
\begin{minipage}[t]{81mm}
\epsfig{bbllx=25,bblly=88,bburx=531,bbury=445,
file=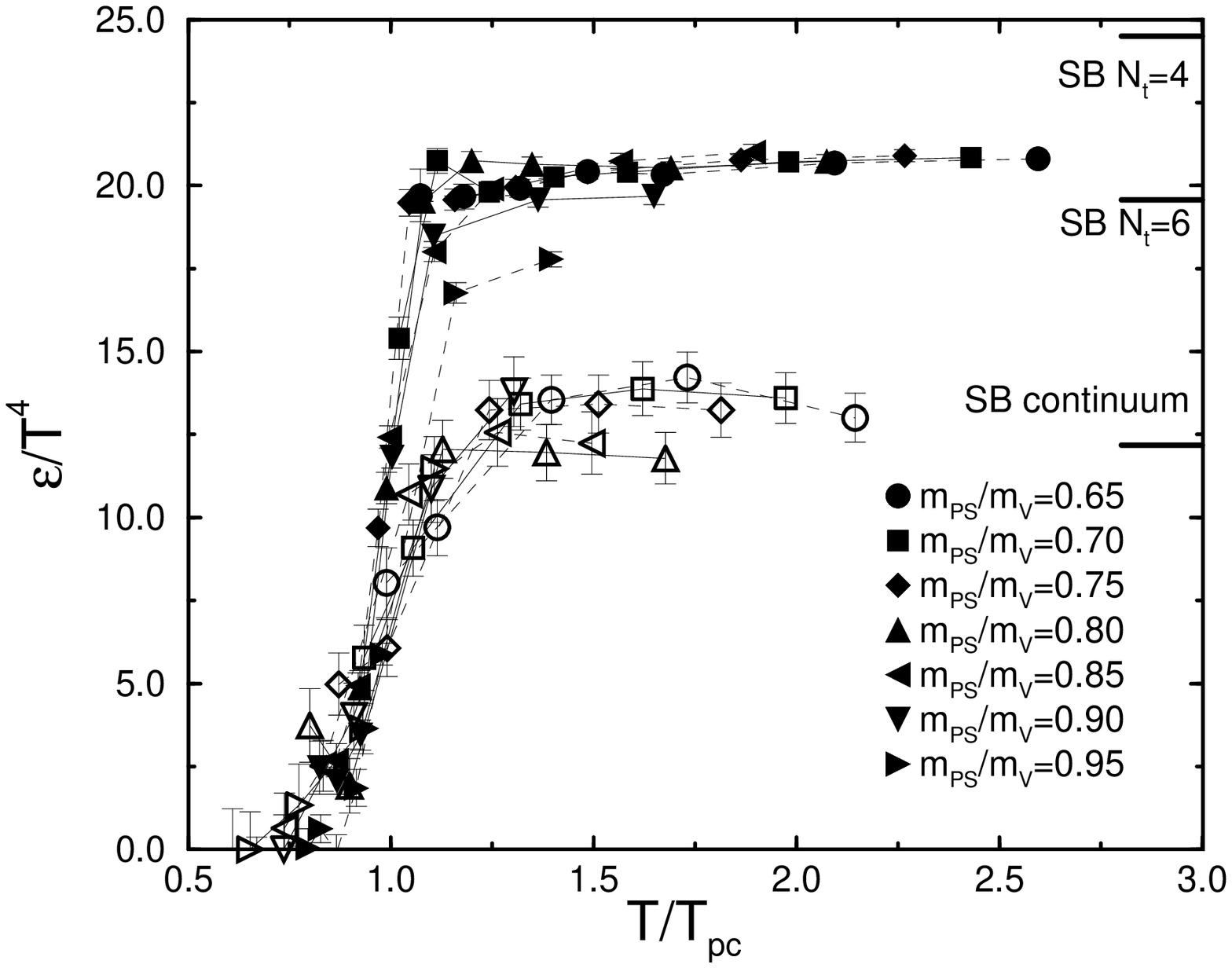,width=81mm}
\end{minipage}
\vskip -0.3truecm
\caption{The energy density in QCD. The left (right) figure shows results
from a calculation with improved staggered (Wilson) fermions on lattices
with temporal extent $N_\tau=4$ ($N_\tau=4,~6$). Arrows in the left 
figure show the ideal gas values $\epsilon_{SB}$ as given by Eq.~3. 
}
\label{fig:energy_nf}
\end{figure*}

Unfortunately, Wilson actions with similarly good high temperature
behaviour have not been constructed so far. The Clover action does
not improve the ideal gas behaviour, \ie it has the same infinite
temperature limit as the Wilson action. Consequently one observes
an overshooting of the ideal gas limit at high temperature which
reflects the cut-off effects in the unimproved fermion sector
\cite{Ali01b}. These cut-off effects are, however, unimportant in
the vicinity of the phase transition where correlation lengths become large.
It thus makes sense to compare results obtained with different actions
in this regime. In Figure~\ref{fig:energy_nf} we show recent results
for the energy density obtained with improved staggered\footnote{This
figure for staggered fermions is based
on data from Ref~\cite{Kar00}. Here a contribution to $\epsilon/T^4$ which 
is proportional to the
bare quark mass and vanishes in the chiral limit is not taken into 
account.} and Wilson
\cite{Ali01b} fermions.
We note that these calculations yield consistent estimates for the 
energy density at $T_c$
\begin{equation}
\epsilon_c \simeq (6 \pm 2) T_c^4 \quad .
\end{equation}
This estimate also is consistent with results obtained for the energy
density from calculations with a standard staggered fermion action
\cite{milc}.

\section{Heavy quark free energy and screening below $T_c$}

For the discussion of medium effects on 
heavy quark bound states in the QGP it is of great importance to
calculate thermal modifications of the heavy quark colour singlet
potential. Unfortunately this is not directly accessible to lattice
calculations, in particular not as a gauge invariant observable.
At finite temperature one generally calculates the heavy quark free
energy \cite{McL} from correlation functions of Polyakov loops $L(\vec n)$, 
\begin{equation}
{\rm e}^{-V(r,T)/T} \equiv \langle {\rm Tr}L(\vec 0 ) 
{\rm Tr}L^\dagger (\vec n) \rangle \quad , \quad rT = |\vec n | N_\tau \; ,
\label{polyakov}
\end{equation} 
where $N_\tau$ denotes the temporal extent of the lattice.
At finite temperature the static $q\bar{q}$-pair can be in a singlet or octet
state and the above correlation function thus only provides a thermal
average over singlet and octet free energies,
\begin{equation}
{\rm e}^{-V(r,T)/T} \equiv {1 \over 9} {\rm e}^{-V_1(r,T)/T}\; + \;
{8 \over 9} {\rm e}^{-V_8(r,T)/T} \quad . 
\label{free18}
\end{equation} 
Although $V_1 \equiv -8\; V_8$ in leading order perturbation theory, we do
in general have to deal with both contributions to $V(r,T)$ and 
cannot directly extract the singlet component. Nonetheless the latter 
will dominate at short distances as well as at low temperature.

\begin{figure}[t]
\vspace{-0.1cm}
\begin{center}
\hspace*{-0.2cm}\epsfig{file=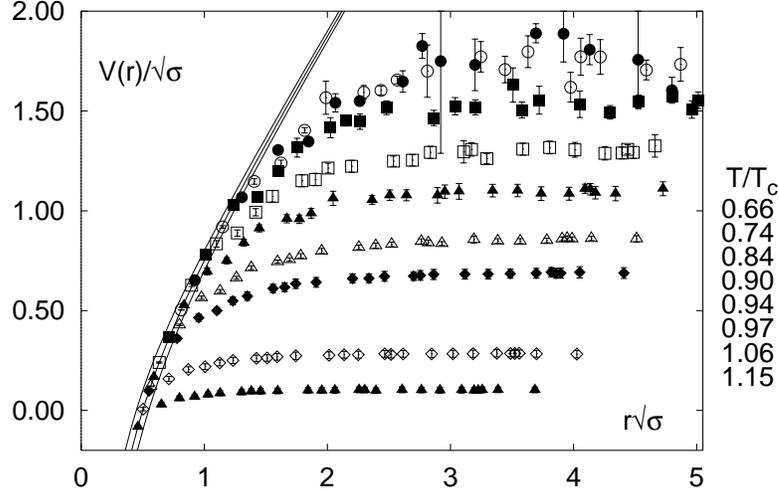,width=100mm}
\end{center}
\vskip -1.0truecm
\caption{Temperature dependence of the heavy quark free energy in
3-flavour QCD \cite{Kar00}. The band of solid curves shows the Cornell 
potential $V(r) =-\alpha /r \; +\; \sigma r$ with 
$\alpha =0.25 \pm 0.05$.
The finite temperature free energies have been normalized to this
potential at the shortest distance available, \ie at $rT = 0.25$.}  
\vskip -0.3truecm
\label{fig:potential}
\end{figure}

The colour averaged heavy quark free energy defined by Eq.~\ref{polyakov} 
thus yields only indirect information for the discussion of thermal
effects on heavy quark bound states. Nonetheless it reflects basic
properties like the strong Debye screening of the $q\bar{q}$-pair above
$T_c$. This has been analyzed in detail in the pure SU(3) gauge
theory \cite{Kaczmarek}. Additional 
screening effects do arise in QCD from the presence of light quarks.   
In addition to quantitative changes in the values of screening
lengths the most significant changes occur in
the long distance behaviour of $V(r,T)$.
While in the pure gauge sector screening of the heavy quark free energy
sets in only above $T_c$, \ie $\displaystyle{\lim_{r \rightarrow \infty}
V(r,T)}$ remains finite only for $T\; \ge \; T_c$, this is the case at all
temperatures as soon as $m_q < \infty$. The temperature dependence of this
screening effect, which arises from the spontaneous creation of 
$q\bar{q}$-pairs in the heat bath, is shown in Figure~\ref{fig:potential}.

As can bee seen the free energy seems to reach a limiting form for 
$T \lsim 0.6\; T_c$. For these temperatures it agrees with the confining
Cornell type potential up to distances $r\sqrt{\sigma}\simeq 1.5$, \ie
$r \simeq 0.8\;$fm. With increasing temperature this point is shifted
to smaller distances. For $T \simeq 0.95\; T_c$ screening sets in already
for $r\sqrt{\sigma}\simeq 0.8$ or $r \simeq 0.4\;$fm. This clearly will
also have an impact on heavy quark bound states. However, to quantify this
we need further information on the short distance properties of the 
heavy quark free energy which should allow to disentangle contributions
from the singlet and octet components.

\section{Thermal Masses}

The analysis of the temperature dependence of hadron properties,
e.g. their masses and widths, is of central importance for our 
understanding of deconfinement and chiral symmetry restoration at
$T_c$.  Thermal modifications of the heavy quark potential which have 
been discussed above influence the spectrum of heavy quark bound states.
Their experimentally observed suppression \cite{jpsi} thus is expected to
be closely linked to the deconfining properties of QCD above $T_c$ 
\cite{Satz}. Changes in the chiral condensate, on the
other hand, influence the light hadron spectrum and may result in 
experimental signatures, for instance in the enhanced 
dilepton production observed in heavy ion experiments \cite{lepton}.

In numerical calculations on Euclidean lattices one has
access to thermal Green's functions $\bar{G}_H(\tau, \vec{r})$
in fixed quantum number channels $H$, to which in particular at high 
temperature many excited states contribute. As the temporal
direction of the Euclidean lattice is rather short at finite 
temperature one usually has restricted numerical investigations to the
analysis of the long distance behaviour of spatial correlations functions,
$\bar{G}_H(\tau, \vec{r}) \sim \exp(-\bar{m}_H |\vec{r}|)$, which 
defines hadronic screening masses $\bar{m}_H$. This indeed gives evidence for
the restoration of chiral symmetry above $T_c$, e.g. one finds that
scalar and pseudo-vector screening lengths become degenerate and also
the difference between screening lengths in scalar and pseudo-scalar 
channels strongly diminishes, which gives indications for a partial 
restoration of the $U_A(1)$ symmetry \cite{Karpisa}.

\begin{figure}[t]
\hspace*{-0.2cm}\epsfig{file=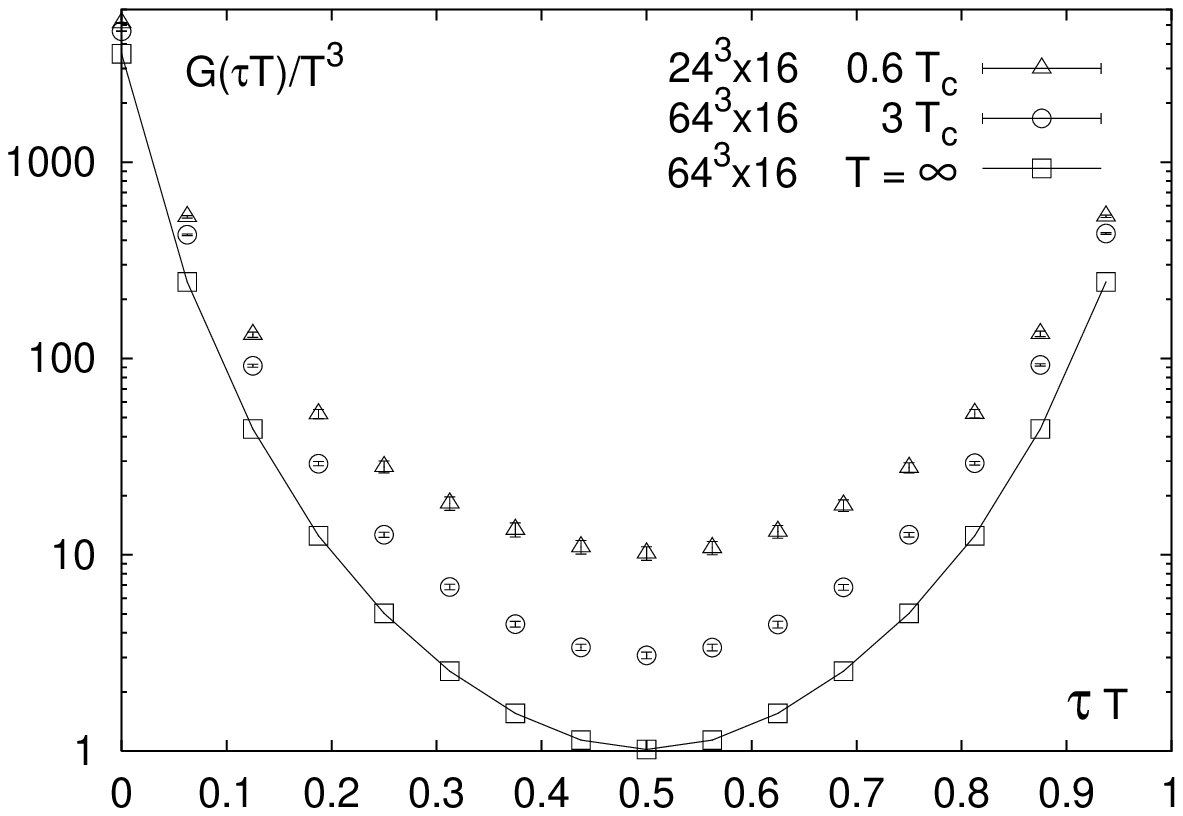,width=82mm}
\hspace*{-0.2cm}\epsfig{file=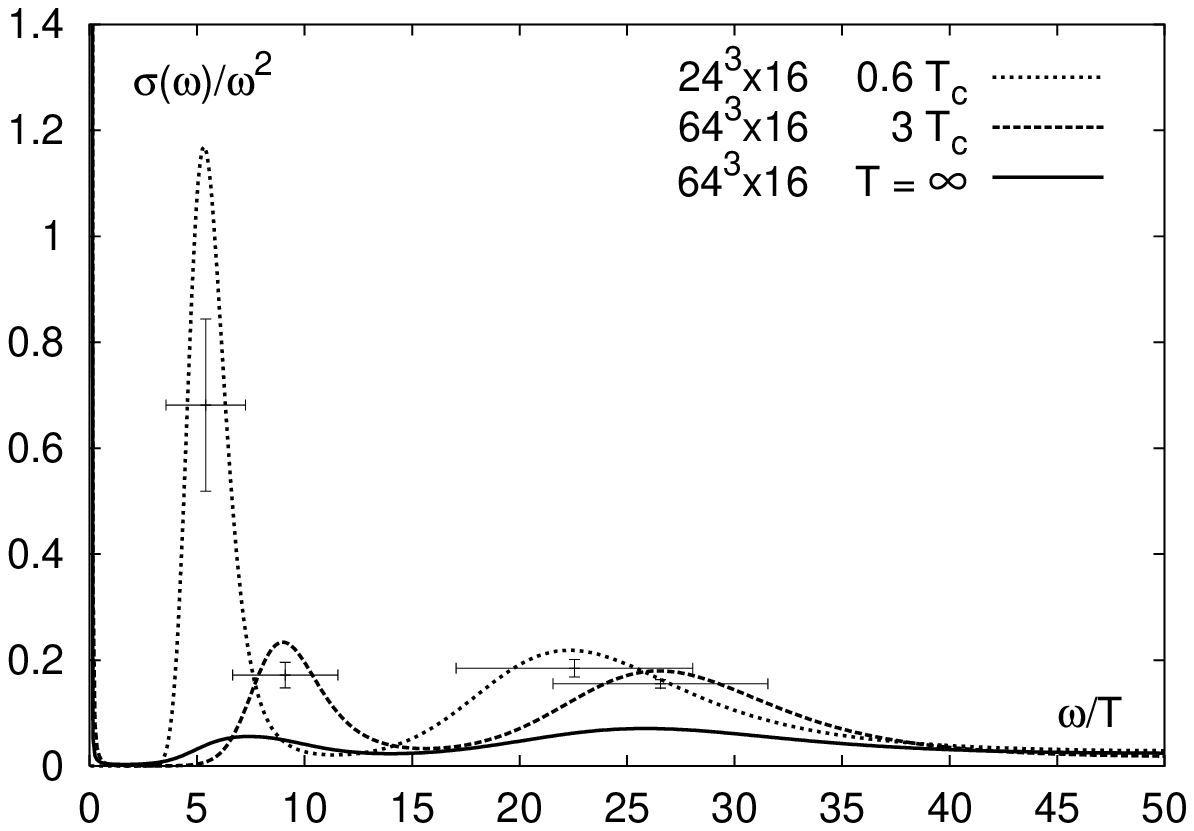,width=82mm}
\vskip -0.7truecm
\caption{A first attempt to reconstruct the thermal pseudo-scalar
spectral function (right) from the temporal correlation function (left)
calculated within the static approximation of QCD at $T=0.6T_c$ 
and $3T_c$ on lattices of size $24^3\times 16$ and $64^3 \times 16$, 
respectively \cite{prelim}.}
\vskip -0.4truecm
\label{fig:spectral}
\end{figure}

In order to get information on the $T$-dependence of pole masses
and their widths one has to analyze the structure of temporal 
correlation functions. The information on hadron masses and quasi-particle
excitations is then encoded in the spectral function $\sigma_H(\omega,\vec{p})$,
\begin{equation}
G _H (\tau, \vec{p}) = \int {d}^3r\; \exp{(i\; \vec{p}\; \vec{r} )} \;
\bar{G}_H (\tau,\vec{r}) = \int_0^\infty d\omega \; \sigma_H(\omega,\vec{p})\;
{\cosh (\omega(\tau-\beta/2)) \over \sinh (\omega\beta/2)} \quad .
\label{tempcor}
\end{equation}
At finite temperature the temporal correlation function usually  
is determined only at a small number of lattice grid points as the
temperature is related to the finite extent of the lattice in this
direction, $N_\tau = 1/(aT)$. A way out may be the use of anisotropic
lattices \cite{For01}. These calculations indeed show large changes in
meson correlators in the vicinity of $T_c$. To what extent the results
suggest that pole masses in light meson channels exist as well defined
states even above $T_c$, however, is difficult to judge solely on the 
basis of standard analysis techniques also used for correlation functions
at zero temperature. At high temperature refined statistical techniques
like the maximum entropy method \cite{Jar96} may be of help. It has recently
been shown that this does allow to extract directly the spectral
functions of hadron correlators \cite{Asa99,Hat01} at $T=0$ as well as
high temperature \cite{wetzorke}. A first attempt to 
use this approach for the determination of the pseudo-scalar spectral 
function below and above $T_c$ is shown in Figure~\ref{fig:spectral}.
This clearly reflects the drastic, qualitative change which occurs in the  
pseudo-scalar channel when one crosses $T_c$. The Goldstone pole which 
is present in the spectral function below $T_c$ disappears in the plasma
phase. Of course, this has to be analyzed more quantitatively in the future.

\section{Conclusions}

We have focused in this review on the calculation of basic thermodynamic 
quantities which are of immediate interest to experimental searches for 
the Quark Gluon Plasma. 

Calculations of the QCD transition temperature which are based on 
different discretization schemes do provide a rather consistent
picture. Extrapolations to the chiral limit are in good agreement
and suggest a transition temperature of about 175~MeV in 2-flavour
QCD. The numerical results obtained so far suggest a rather weak quark 
mass and flavour dependence of $T_c$. This also holds true for
the critical energy density, which for 2-flavour QCD is estimated 
to be $\epsilon_c \simeq 700$~MeV and thus is of similar magnitude as that
found already in the pure gauge sector. 
However, the error on $\epsilon_c$ still is about 50\% and 
mainly is due to the current uncertainty on $T_c$ which is estimated to be
about 10\%.

An analysis of the heavy quark free energy below $T_c$ shows that
screening effects  increase significantly already below $T_c$. 
While the spontaneous creation of $q\bar{q}$-pairs leads to 
screening of the heavy quark potential at $r\simeq 0.8$~fm at low
temperature this starts already at $r\simeq 0.4$~fm for $T\simeq 0.95 T_c$. 
  
We do have reached some understanding of thermal effects on hadron
properties. In particular, modifications of the light meson spectrum
due to flavour and approximate $U_A(1)$ symmetry restoration have
been established. Indications for medium effects on pole
masses have been found in an analysis of spectral
functions.  However, at present these calculations are not detailed enough 
to be confronted with experimental data. 

Of course, there are many other important issues which have to be
addressed in the future. Even at vanishing baryon number density
we do not yet have a satisfactory understanding of the critical
behaviour of 2-flavour QCD in the chiral limit and 
the physically realized situation of QCD with two light, nearly
massless quarks and a heavier strange quark has barely been
analyzed.     
Moreover, the entire phase diagram at non-zero baryon number 
density is largely unexplored. An interesting phase structure
is predicted in this case which currently
is not accessible to lattice calculations.

\medskip
\noindent
{\bf Acknowledgements:} I would like to thank 
N. Christ for communication on recent thermodynamic results with
domain wall fermions.

\end{document}